\documentclass{article}
\usepackage{spconf,amsmath,graphicx,bm,amssymb,amsthm,dsfont}
\usepackage{cite}
\usepackage{hyperref}
\hypersetup{
	colorlinks=true,
	linkcolor=blue,
	filecolor=magenta,
	urlcolor=cyan,
	citecolor=blue
}

\newcommand{\argmax}{\arg\max}

\newcommand{\norm}[1]{\left\|#1\right\|}
\newcommand{\normm}[1]{\big\|#1\big\|}
\newcommand{\normv}[1]{\big\|#1\big\|}

\newtheorem{theorem}{Theorem}%[section]
\newtheorem{lemma}{Lemma}%[section]
\newtheorem{assumption}{Assumption}

\title{
	Scaling Law Analysis for Covariance Based Activity Detection in Cooperative Multi-Cell Massive MIMO
}
\name{Ziyue Wang$^{\star,\S}$, Ya-Feng Liu$^{\star}$, Zhaorui Wang$^{\dag}$, and Wei Yu$^{\ddag}$}

\address{$^{\star}$LSEC, ICMSEC, AMSS, Chinese Academy of Sciences, Beijing, China\\%[2pt]
    $^{\S}$School of Mathematical Sciences, University of Chinese Academy of Sciences, Beijing, China\\%[2pt]
    $^{\dag}$SSE and FNii, The Chinese University of Hong Kong, Shenzhen, China\\%[2pt]
    $^{\ddag}$Department of Electrical and Computer Engineering, University of Toronto, Toronto, Canada\\%[2pt]
  Email: \{ziyuewang, yafliu\}@lsec.cc.ac.cn, wangzhaorui@cuhk.edu.cn, weiyu@ece.utoronto.ca
}
\begin{document}
\ninept

\setlength{\abovedisplayskip}{0.08cm}
\setlength{\belowdisplayskip}{0.08cm}
\maketitle
%
%\wuhao {
\begin{abstract}
This paper studies the covariance based activity detection problem in a multi-cell massive multiple-input multiple-output (MIMO) system, where the active devices transmit their signature sequences to multiple base stations (BSs), and the BSs cooperatively detect the active devices based on the received signals.
The scaling law of covariance based activity detection in the \emph{single-cell} scenario has been thoroughly analyzed in the literature. This paper aims to analyze the scaling law of covariance based activity detection in the \emph{multi-cell} massive MIMO system. In particular, this paper shows a quadratic scaling law in the multi-cell system under the assumption that the exponent in the classical path-loss model is greater than $2,$ 
which demonstrates that in the multi-cell MIMO system the maximum number of active devices that can be correctly detected in each cell increases quadratically with the length of the signature sequence and decreases logarithmically with the number of cells (as the number of antennas tends to infinity).
This paper also characterizes the distribution of the estimation error in the multi-cell scenario.
\end{abstract}
\begin{keywords}
Cooperative activity detection, massive random access, multi-cell massive MIMO, scaling law analysis. %Asymptotic performance analysis, 
\end{keywords}
\section{Introduction}
\label{sec:intro}

Massive machine-type communication (mMTC) is an important application scenario in the fifth-generation (5G) and beyond cellular systems \cite{bockelmann2016massive}. A main challenge in mMTC is massive random access, in which a large number of devices are connected to the network, but the device activities are sporadic \cite{chen2021massive}.
Activity detection can be accomplished in the first stage of the three-phase protocol \cite{kang2021minimum,kang2022scheduling},
during which the active devices transmit their unique preassigned non-orthogonal signature sequences, and the network then identifies the active devices by detecting which sequences are transmitted based on the received signals at the base-stations (BSs) \cite{liu2018sparse}. The non-orthogonality of the signature sequences inevitably causes both intra-cell and inter-cell interference that complicates the task of device activity detection.

The device activity detection problem can be formulated as a compressed sensing (CS) problem, in which device activities can be estimated jointly with the instantaneous channel state information (CSI) by sparse recovery methods \cite{senel2018grant,liu2018massive,chen2018sparse,liu2021efficient}.
%When the CSI is not needed and
If the CSI is not needed, 
the activity detection problem can be formulated as a maximum likelihood estimation (MLE) problem \cite{haghighatshoar2018improved}, which is often called the covariance based approach because the problem depends only on the sample covariance matrix of the received signals.
The covariance based approach was first proposed in the pioneering work \cite{haghighatshoar2018improved} for the single-cell scenario and then was extended to the multi-cell scenario in \cite{chen2021sparse,ganesan2021clustering}.
The covariance based approach was also used for many related data/activity detection problems in different scenarios; see \cite{fengler2021non,chen2022phase,chen2019covariance,shao2020cooperative,wang2022covariance,jiang2022ml,liu2022mle_spawc,liu2022mle_icc,jia2021statistical} for more details.
State-of-the-art algorithms for solving the MLE problem includes coordinate descent (CD) \cite{haghighatshoar2018improved} and  accelerated versions \cite{wang2021accelerating,dong2022faster}, gradient descent \cite{wang2021efficient}, expectation minimization \cite{wipf2007empirical}, SPICE \cite{yang2018sparse}, and penalty based algorithms \cite{lin2022sparsity,li2022asynchronous}. 
It has been shown in \cite{chen2022phase,chen2019covariance} that the detection performance of the covariance based approach is generally much better than that of the CS based approach, especially in the massive multiple-input multiple-output (MIMO) system.
In this paper, we focus on the massive MIMO system and the covariance based approach.

In addition to algorithmic development, an important line of research for covariance based activity detection is to analyze its scaling law,
i.e., characterize the feasible set of system parameters under which this covariance based approach is able to successfully recover the device activities in the massive MIMO system.
Along this line, the scaling law of the covariance based approach has been analyzed in \cite{fengler2021non,chen2022phase} in the single-cell scenario.
More specifically, \cite{fengler2021non} showed that the maximum number of active devices that can be correctly detected scales as $K=\mathcal{O}(L^2/\log^2(N/L^2))$ by solving the restricted MLE problem,
where $N$ is the number of potential devices and $L$ is the signature sequence length.
The work in \cite{chen2022phase} further showed that the above scaling law also holds true for the more practical unrestricted MLE model.
As far as we know, \cite{chen2021sparse} is the first and the only work on the phase transition analysis of the unrestricted MLE model/the covariance based approach in multi-cell massive MIMO. In particular, it is conjectured and verified by simulation in \cite{chen2021sparse} that the scaling law of the covariance based approach in the multi-cell scenario is approximately the same as the single-cell scenario. 

In this paper, we study the scaling law of covariance based activity detection in the cooperative \emph{multi-cell} massive MIMO system.
In particular, we consider a cloud radio access network (C-RAN) architecture, which consists of $B$ cells connected to a central unit (CU) via fronthaul links and $N$ potential devices per cell.
We assume that all large-scale fading coefficients are known and they satisfy the path-loss model in \cite{rappaport1996wireless} with the path-loss exponent being greater than $2.$
Under the above assumptions, we show that, as the number of antennas tends to infinity, the maximum number of active devices in each cell that can be detected at the CU is $K=\mathcal{O}(L^2/\log^2(BN/L^2)),$ which is approximately the same as that of the single-cell scenario \cite{fengler2021non,chen2022phase}.
The above scaling law shows that the covariance based approach can correctly detect almost as many active devices in each cell in the multi-cell scenario as it does in the single-cell scenario, and the inter-cell interference is not a limiting factor of the detection performance because $B$ affects $K$ only through $\log B.$
This result clarifies the condition under which the conjecture in \cite{chen2021sparse} holds true.
Moreover, we also characterize the distribution of the estimation error, and show that it can be numerically computed by solving quadratic programming (QP).

\section{System Model and Problem Formulation}

\subsection{System Model}

Consider an uplink multi-cell massive MIMO system consisting of $B$ cells.
Each cell contains one BS equipped with $M$ antennas and $N$ single-antenna devices.
We assume that a C-RAN architecture is used for inter-cell interference mitigation, in which all $B$ BSs are connected to a CU via fronthaul links such that the signals received at the BSs can be collected and jointly processed at the CU.
Assume that during any coherence interval, only $K\ll N$ devices are active in each cell.
For device identification, each device $n$ in cell $j$ is preassigned a unique signature sequence $\mathbf{s}_{jn}\in\mathbb{C}^{L}$ with $L$ being the sequence length.
Let $a_{jn}$ be a binary variable with $a_{jn}=1$ for active and $a_{jn}=0$ for inactive devices.
The channel between device $n$ in cell $j$ and BS $b$ is denoted as $\sqrt{g_{bjn}}\mathbf{h}_{bjn},$ where $g_{bjn} \geq 0$ is the large-scale fading coefficient depending on path-loss and shadowing, and $\mathbf{h}_{bjn}\in\mathbb{C}^{M}$ is the i.i.d. Rayleigh fading component following $\mathcal{CN}(\mathbf{0},\mathbf{I}).$

In the uplink pilot stage, all active devices synchronously transmit their preassigned signature sequences to BSs as random access requests.
Then, the received signal at BS $b$ can be expressed as
\begin{align}\label{eq.sys}
	\mathbf{Y}_b&=\sum_{n=1}^{N}a_{bn}\mathbf{s}_{bn}g_{bbn}^{\frac{1}{2}}\mathbf{h}_{bbn}^T  + \sum_{j\neq b} \sum_{n=1}^N a_{jn}\mathbf{s}_{jn}g_{bjn}^{\frac{1}{2}}\mathbf{h}_{bjn}^T+\mathbf{W}_b\nonumber\\
	&=\mathbf{S}_b\mathbf{A}_b\mathbf{G}^{\frac{1}{2}}_{bb}\mathbf{H}_{bb}+\sum_{j\neq b}\mathbf{S}_j\mathbf{A}_j\mathbf{G}^{\frac{1}{2}}_{bj}\mathbf{H}_{bj}+\mathbf{W}_b,
\end{align}
where $\mathbf{S}_j=[\mathbf{s}_{j1},\ldots,\mathbf{s}_{jN}]\in\mathbb{C}^{L\times N}$ is the signature sequence matrix of the devices in cell $j,$ $\mathbf{A}_{j}=\operatorname{diag}(a_{j1},\ldots,a_{jN})$ is a diagonal matrix that indicates the activity of the devices in cell $j,$ $\mathbf{G}_{bj}=\operatorname{diag}(g_{bj1},\ldots,g_{bjN})$ contains the large-scale fading components between the devices in cell $j$ and BS $b,$ $\mathbf{H}_{bj}=[\mathbf{h}_{bj1},\ldots,\mathbf{h}_{bjN}]^T\in\mathbb{C}^{N\times M}$ is the Rayleigh fading channel between the devices in cell $j$ and BS $b,$ and $\mathbf{W}_b$ is the additive Gaussian noise that follows $\mathcal{CN}(\mathbf{0},\sigma_w^2\mathbf{I}),$ where $\sigma_w^2$ is the variance of the background noise normalized by the transmit power. %for notational simplicity.

For notational simplicity, 
let $\mathbf{A} = \operatorname{diag}( \mathbf{A}_1, \ldots, \mathbf{A}_B ) \in \mathbb{R}^{BN \times BN}$ be a diagonal matrix that indicates the activity of all devices,
and let $\mathbf{a}\in\mathbb{R}^{BN}$ denote its diagonal entries. %denotes the  of $\mathbf{A}.$
Let $\mathbf{S}=[\mathbf{S}_{1},\ldots,\mathbf{S}_{B}]\in\mathbb{C}^{L\times BN}$ denote the signature matrix of all devices, and
let $\mathbf{G}_b=\operatorname{diag}(\mathbf{G}_{b1},\ldots,\mathbf{G}_{bB})\in\mathbb{R}^{BN\times BN}$ denote the matrix containing large-scale fading components between all devices and BS $b.$

\subsection{Problem Formulation}

The activity detection problem is to detect the active devices from the received signals $\mathbf{Y}_b,\, b=1,\ldots,B.$
In this paper, we assume the large-scale fading coefficients are known, i.e., the matrices $\mathbf{G}_{b}$ for all $b$ are known at the BSs.
In this case, the activity detection problem is equivalent to estimating the activity indicator vector $\mathbf{a}.$

Notice that for each BS $b,$ the Rayleigh fading components and noises are both i.i.d.\ Gaussian over the antennas.
Thus, for a given $\mathbf{a},$ the columns of received signal $\mathbf{Y}_{b}$ in \eqref{eq.sys} denoted by $\mathbf{y}_{bm},\,m=1,\ldots,M$ are i.i.d. Gaussian vectors,
that is $\mathbf{y}_{bm}\sim\mathcal{CN}(\mathbf{0},\boldsymbol\Sigma_b),$ where the covariance matrix $\boldsymbol\Sigma_b$ is given by
\vspace{2pt}
\begin{equation}%\label{eq.sys.multi.cell}
	\boldsymbol\Sigma_b = \frac{1}{M}\mathbb{E}\left[\mathbf{Y}_b\mathbf{Y}_b^H\right]=\mathbf{S} \mathbf{G}_b \mathbf{A} \mathbf{S}^H + \sigma_w^2 \mathbf{I}.
	\vspace{2pt}
\end{equation}
Since the received signals $\mathbf{Y}_b,\,b = 1,\ldots,B$ are independent due to the i.i.d. Rayleigh fading channels, the likelihood function $p(\mathbf{Y}_1,\ldots,\mathbf{Y}_B|~\mathbf{a}) = \Pi_{b=1}^{B}p(\mathbf{Y}_b|~\mathbf{a}).$ 
Hence the maximum likelihood estimation (MLE) problem is equivalent to the minimization of $- \frac{1}{M} \Sigma_{b=1}^{B} \log p(\mathbf{Y}_b|~\mathbf{a}),$ which can be formulated as \cite{chen2021sparse}
\begin{subequations}\label{eq.prob1.multi}
	%\vspace{-0pt}
	\begin{alignat}{2}\label{eq.prob1.2.multi}
		&\underset{\mathbf{a}}{\operatorname{minimize}}    &\quad& \sum_{b=1}^B\left(\log\left|\boldsymbol\Sigma_b\right|+ \operatorname{tr}\left(\boldsymbol\Sigma_b^{-1}\widehat{\boldsymbol\Sigma}_b\right)\right)\\
		&\operatorname{subject\,to} &      &a_{bn} \in [0,1], \,\forall \, b,n,
		\label{eq.prob1.3.multi}
		%\vspace{5pt}
	\end{alignat}
\end{subequations}
where $\widehat {\boldsymbol\Sigma}_b = \mathbf{Y}_b\mathbf{Y}_b^H / M$ is the sample covariance matrix of the received signals.

In this paper, we are interested in characterizing the detection performance limit of problem \eqref{eq.prob1.multi} as the number of antennas $M$ tends to infinity.
In particular, we are interested in answering the following two questions in the next section:
(i) given the system parameters $L, B,$ and $N,$ how many active devices can be correctly detected via solving the MLE problem \eqref{eq.prob1.multi} as $M\rightarrow \infty?$
(ii) what is the asymptotic distribution of the MLE error?

\section{Main Results}

In this section, we present the main results of this paper, i.e., the scaling law and the estimation error distribution of the solution to the MLE problem \eqref{eq.prob1.multi}.
Both of the above results are based on the consistency result of the MLE problem \eqref{eq.prob1.multi} derived in \cite{chen2021sparse} in a multi-cell setup.
\begin{lemma}[Consistency of MLE (Theorem 3 in \cite{chen2021sparse})]\label{lemma:consistency}
	Consider the MLE problem \eqref{eq.prob1.multi} with a given signature sequence matrix $\mathbf{S},$ large-scale fading component matrices $\{\mathbf{G}_b\},$ and noise variance $\sigma_w^2.$ 
	Let $\widetilde{\mathbf{S}}$ be the matrix whose columns are the Kronecker product of $\mathbf{s}_{bn}^*$ and $ \mathbf{s}_{bn}$ (where $(\cdot)^*$ is the conjugate operation), i.e.,
	\vspace{3pt}
	\begin{equation}\label{eq:s-tilde}
		\widetilde{\mathbf{S}}=[ \mathbf{s}_{11}^*\otimes \mathbf{s}_{11},\ldots ,\mathbf{s}_{BN}^*\otimes \mathbf{s}_{BN} ]\in\mathbb{C}^{L^2\times BN}.
		\vspace{3pt}
	\end{equation}
	Let $\hat{\mathbf{a}}^{(M)}$ be the solution to \eqref{eq.prob1.multi} when the number of antennas $M$ is given, and let $\mathbf{a}^{\circ}$ be the true activity indicator vector whose $B(N-K)$ zero entries are indexed by $ \mathcal{I},$ i.e., $ \mathcal{I}\triangleq\{i\mid a_i^{\circ}=0\}.$
	%where $a_i^{\circ}$ being the $i$-th entry of $\mathbf{a}^{\circ}.$
	Define\vspace{3pt}
	\begin{align}
		\mathcal{N}&\triangleq\{\mathbf{x}\in \mathbb{R}^{BN}\mid \widetilde{\mathbf{S}}\mathbf{G}_b\mathbf{x}= \mathbf{0}, \, \forall\, b\},\label{eq:subspace} \\
		\mathcal{C}&\triangleq\{\mathbf{x}\in\mathbb{R}^{BN}\mid x_i\geq 0~\text{if}~i\in \mathcal{I},\, x_i\leq 0~\text{if}~i \notin \mathcal{I}\},\label{eq:cone}
	\end{align}
	then a necessary and sufficient condition for $\hat{\mathbf{a}}^{(M)}\to \mathbf{a}^{\circ}$ as $M\to\infty$ is that the intersection of $\mathcal{N}$ and $\mathcal{C}$ is the zero vector, i.e., $\mathcal{N}\cap\mathcal{C}=\{\mathbf{0}\}.$
\end{lemma}

\subsection{Scaling Law Analysis}

One of the main contributions of this paper is to characterize the scaling law of the MLE problem \eqref{eq.prob1.multi}, i.e., the feasible set of the system parameters under which the condition $\mathcal{N}\cap\mathcal{C}=\{\mathbf{0}\}$ in Lemma~\ref{lemma:consistency} holds true.
It turns out that the generation of the signature sequence matrix and the large-scale fading coefficients play vital roles in the scaling law analysis.
Next, we specify the assumptions on the signature sequence matrix and the large-scale fading coefficients as well as the properties under the corresponding assumptions.

\begin{assumption}\label{assu:sequence}
	The columns of the signature sequence matrix $\mathbf{S}$ are uniformly drawn from the sphere of radius $\sqrt{L}$ in an i.i.d. fashion.
\end{assumption}
The following lemma is a direct corollary of \cite[Theorem 2]{fengler2021non}.
\begin{lemma}\label{lemma:nsp}
	Suppose that Assumption~\ref{assu:sequence} holds true. Then, for any given parameter $\bar{\rho}\in (0,1),$ there exist some constants $c_1$ and $c_2$ depending only on $\bar{\rho}$ such that if
	$
	K \le c_1L^2/ \log^2(eBN/L^2),
	$
	then with probability at least $1 - \exp(-c_2L),$ the matrix $\widetilde{\mathbf{S}}$ defined in \eqref{eq:s-tilde} has the stable null space property (NSP) of order $K$ with parameters $\rho \in (0,\bar{\rho}).$ %$0 < \rho < \bar{\rho}.$
	More precisely, for any $\mathbf{v} \in \mathbb{R}^{BN}$ that satisfies $\widetilde{\mathbf{S}} \mathbf{v} = \mathbf{0},$ the following inequality \vspace{1pt}
	\begin{equation}\label{eq:nsp}
		\norm{\mathbf{v}_{\mathcal{K}}}_1 \leq \rho\norm{\mathbf{v}_{\mathcal{K}^c}}_1 \vspace{1pt}
	\end{equation}
	holds for any index set $\mathcal{K}\subseteq\{1,2,\ldots,BN\}$ with $|\mathcal{K}|\leq K,$ where $\mathbf{v}_{\mathcal{K}}$ is a sub-vector of $\mathbf{v}$ with entries from $\mathcal{K},$ and $\mathcal{K}^{c}$ is the complementary set of $\mathcal{K}$ with respect to $\{1,2,\ldots,BN\}.$
\end{lemma}

\begin{assumption}\label{assu:cell}
	The multi-cell system consists of $B$ hexagonal cells with radius $R.$
	The BSs are in the center of the corresponding cells.
	In this system, the large-scale fading components decrease exponentially with distance \cite{rappaport1996wireless}, i.e.,\vspace{1pt}
	\begin{equation}\label{eq:gamma}
		g_{bjn} = P_0 \left(\frac{d_0}{d_{bjn}}\right)^{\gamma},\vspace{1pt}
	\end{equation}
	where $P_0$ is the received power at the point with distance $d_0$ from the transmitting antenna, $d_{bjn}$ is the BS-device distance between device $n$ in cell $j$ and BS $b,$ and $\gamma$ is the path-loss exponent.
\end{assumption}

\begin{lemma}\label{lemma:gamma}
	Suppose that Assumption~\ref{assu:cell} holds true with $\gamma>2.$
	Then, there exists a constant $C > 0$ depending only on $\gamma, P_0, d_0,$ and $R$ defined in Assumption~\ref{assu:cell}, such that for each BS $b,$ the large-scale fading coefficients satisfy
	\vspace{2pt}
	\begin{equation}\label{eq:less-C}
		\sum_{j=1,\,j\neq b}^{B} \Big(\max_{1\le n \le N} g_{bjn} \Big) \le C.
		\vspace{-2pt}
	\end{equation}
\end{lemma}
Lemma~\ref{lemma:gamma} shows that as long as the path-loss exponent $\gamma>2$ in Assumption~\ref{assu:cell}, then the summation in the left-hand side of \eqref{eq:less-C} is upper bounded by a constant $C$ which is independent of $B.$
Instead of rigorously proving the lemma, we give its engineering interpretation as follows. For a particular BS $b,$ since most of the interfering cells are far away from it, the distance $d_{bjn}$'s for all devices in these cells will be large.
In fact, the term $\max_{n} g_{bjn}$ in the summation in the left-hand side of \eqref{eq:less-C} will converge to zero sufficiently fast such that it is negligible (due to Eq. \eqref{eq:gamma}).
It turns out that $\gamma>2$ is a sufficient condition for \eqref{eq:less-C} and in fact $\gamma>2$ holds true for most channel models and application scenarios \cite{rappaport1996wireless}.
Finally, it is worthwhile remarking here that the hexagonal structure is not essential in Lemma~\ref{lemma:gamma}. Similar results can be derived for other structured systems such as squared multi-cell systems.

Now we are ready to present one of the main results of this paper, which provides an analytic scaling law by establishing a sufficient condition for $\mathcal{N}\cap \mathcal{C}=\{\mathbf{0}\}.$
\begin{theorem}\label{theorem:scaling-law}
	Under Assumption~\ref{assu:sequence} and Assumption~\ref{assu:cell} with $\gamma>2,$ then there exist constants $c_1,c_2 > 0$ independent of system parameters $K, L, N,$ and $B,$ such that if \vspace{2pt}
	\begin{equation}\label{eq:scaling-law}
		K \le c_1 L^2 / \log^2(eBN/L^2),\vspace{2pt}
	\end{equation}
	then the condition $\mathcal{N}\cap \mathcal{C}=\{\mathbf{0}\}$ in Lemma~\ref{lemma:consistency} holds with probability at least $1-\exp(-c_2L).$
\end{theorem}

A few remarks on Theorem~\ref{theorem:scaling-law} are in order.
First, Theorem~\ref{theorem:scaling-law} states that $\hat{\mathbf{a}}^{(M)}\to \mathbf{a}^{\circ}$ as $M\to\infty$ holds with overwhelmingly high probability under the scaling law \eqref{eq:scaling-law}.
Therefore, with a sufficiently large $M,$ the maximum number of active devices that can be correctly detected by solving the MLE problem \eqref{eq.prob1.multi} is in the order of $L^2$ shown in \eqref{eq:scaling-law}.
Second, the scaling law in \eqref{eq:scaling-law} in the multi-cell scenario is approximately the same as the single-cell scenario \cite{fengler2021non,chen2022phase}, 
which implies that solving the MLE problem \eqref{eq.prob1.multi} can detect almost as many active devices in each cell in the multi-cell scenario as it does in the single-cell scenario.
It also shows that the inter-cell interference is not a limiting factor of the detection performance because $B$ affects $K$ only through $\log B.$ 

\textit{Proof of Theorem~\ref{theorem:scaling-law}:}
Theorem~\ref{theorem:scaling-law} can be proved by contradiction. The outline of the proof is as follows: first we use the stable NSP of the matrix $\widetilde{\mathbf{S}}$ (in Lemma~\ref{lemma:nsp}) to derive some useful inequalities; then we further look into the derived inequalities by carefully exploiting properties of sets in \eqref{eq:subspace} and \eqref{eq:cone}; finally we use the property of the path-loss model (in Lemma~\ref{lemma:gamma}) to derive the contradiction.

Now we prove Theorem~\ref{theorem:scaling-law} by contradiction.
First, suppose that there is a non-zero vector $\mathbf{x} \in \mathcal{N}\cap\mathcal{C}.$
Then we repeatedly use Lemma~\ref{lemma:nsp} to obtain $B$ inequalities.
Define $\mathcal{A}_b$ as an index set of $\mathbf{a}^{\circ},$ corresponding to all active devices in cell $b,$ i.e., $|\mathcal{A}_b| = K$ and
\vspace{1pt}
\begin{equation}
	\mathcal{A}_b = \{ i \mid a_i^{\circ}=1,\, (b-1)N+1 \le i \le bN \}, \, 1 \le b \le B.\vspace{1pt}
\end{equation}
Define $D\triangleq P_0\left(\frac{d_0}{R}\right)^{\gamma}.$ Then it follows from \eqref{eq:gamma} that\vspace{1pt}
\begin{equation}\label{eq:large-D}
	\min_{1 \le n \le N} g_{bbn} \ge D, \,   1 \le b \le B.%\forall b.
	\vspace{1pt}
\end{equation}
Set $\bar{\rho} = \frac{D}{D+2C}$ in Lemma~\ref{lemma:nsp}, where $C$ is defined in \eqref{eq:less-C}. Recall the definition of $\mathcal{N}$ in \eqref{eq:subspace}, for each cell $b,$ setting the index set $\mathcal{K} = \mathcal{A}_b$ and $\mathbf{v} = \mathbf{G}_b\mathbf{x}$ in \eqref{eq:nsp}, we get $B$ inequalities:
\vspace{2pt}
\begin{equation}\label{eq:nsp-main}
	\normm{\left[ \mathbf{G}_b\mathbf{x} \right]_{\mathcal{A}_b}}_1 \leq \rho\normv{\left[ \mathbf{G}_b\mathbf{x} \right]_{\mathcal{A}_b^c}}_1, \, 1 \le b \le B.
	\vspace{1pt}
\end{equation}

Next, we derive an equivalent form of \eqref{eq:nsp-main} by studying its right-hand side.
Notice that $\mathcal{A}_b^c = \left( \cup_{j\neq b} \mathcal{A}_j \right) \cup \mathcal{I}.$ Then we get\vspace{-1pt}
\begin{equation}\label{eq:l1-eq-com-sum}
	\normm{\left[ \mathbf{G}_b\mathbf{x} \right]_{\mathcal{A}_b^c}}_1 = \normm{\left[ \mathbf{G}_b\mathbf{x}\right]_{\mathcal{I}}}_1 + \sum_{j=1,\,j\neq b}^{B} \normm{\left[ \mathbf{G}_b\mathbf{x}\right]_{\mathcal{A}_j}}_1.\vspace{-1pt}
\end{equation}
Since $\mathbf{x} \in \mathcal{C}$ and $\mathbf{G}_b$ is a positive definite diagonal matrix, it follows that the two sub-vectors of $\mathbf{G}_b\mathbf{x}$ with entries from $\mathcal{I}$ and $\mathcal{I}^c$ are non-negative and non-positive, respectively, i.e.,
$\left[ \mathbf{G}_b\mathbf{x}\right]_{\mathcal{I}} \ge \mathbf{0}$ and $\left[ \mathbf{G}_b\mathbf{x}\right]_{\mathcal{I}^c} \le \mathbf{0}.$
Now let $\mathbf{1} \in \mathbb{R}^{BN}$ denote the all-one vector, then
\vspace{1pt}
\begin{equation}\label{eq:minus-l1norm}
	\mathbf{1}^T\mathbf{G}_b\mathbf{x} = \normm{\left[ \mathbf{G}_b\mathbf{x}\right]_{\mathcal{I}}}_1 - \normm{\left[ \mathbf{G}_b\mathbf{x}\right]_{\mathcal{I}^c}}_1.\vspace{2pt}
\end{equation}
Let $\mathbf{u}$ denote the vectorization of the $L \times L$ identity matrix i.e., $\mathbf{u} = \operatorname{vec}\left(\mathbf{I}\right) \in \mathbb{R}^{L^2}.$ Then, for each column of $\widetilde{\mathbf{S}},$ it holds that\vspace{1pt}
\begin{equation}
	\mathbf{u}^T \left( \mathbf{s}_{bn}^*\otimes \mathbf{s}_{bn} \right) = \operatorname{tr} \big( \mathbf{I} \cdot \mathbf{s}_{bn} \mathbf{s}_{bn}^H \big) = \|\mathbf{s}_{bn}\|_2^2.\vspace{1pt}
\end{equation}
Using $\mathbf{x} \in \mathcal{N}$ and the normalization of the sequence $\|\mathbf{s}_{bn}\|_2^2 = L,$ we get
\begin{equation}\label{eq:sum-zero}
	\mathbf{u}^T \widetilde{\mathbf{S}} \mathbf{G}_b \mathbf{x} = L\mathbf{1}^T\mathbf{G}_b\mathbf{x} = 0.
\end{equation}
Combining \eqref{eq:minus-l1norm} and \eqref{eq:sum-zero}, and noting that $\mathcal{I}^c = \cup_{j=1}^{B} \mathcal{A}_j,$ we have\vspace{-1pt}
\begin{equation}\label{eq:l1-eq-sum}
	\normm{\left[ \mathbf{G}_b\mathbf{x}\right]_{\mathcal{I}}}_1 = \normm{\left[ \mathbf{G}_b\mathbf{x}\right]_{\mathcal{I}^c}}_1 = \sum_{j=1}^{B} \normm{\left[ \mathbf{G}_b\mathbf{x}\right]_{\mathcal{A}_j}}_1.\vspace{-1pt}
\end{equation}
By substituting \eqref{eq:l1-eq-sum} into \eqref{eq:l1-eq-com-sum}, we can equivalently rewrite \eqref{eq:nsp-main} as\vspace{-1pt}
\begin{align}\label{eq:ineq-main}
	\normm{\left[ \mathbf{G}_b\mathbf{x} \right]_{\mathcal{A}_b}}_1 \leq \rho & \Big( \normm{\left[ \mathbf{G}_b\mathbf{x} \right]_{\mathcal{A}_b}}_1 + 2\sum_{j=1,\,j\neq b}^{B} \normm{\left[ \mathbf{G}_b\mathbf{x}\right]_{\mathcal{A}_j}}_1 \Big), \nonumber \\
	& \qquad \qquad \qquad \qquad \qquad 1 \le b \le B . \vspace{-1pt}
\end{align}

Finally, we will derive the contradiction by putting all pieces together. Since $\rho<\bar{\rho} <1,$ it follows from \eqref{eq:ineq-main} that\vspace{-1pt}
\begin{equation}\label{eq:reduced-main}
	\normm{\left[ \mathbf{G}_b\mathbf{x} \right]_{\mathcal{A}_b}}_1 \le \frac{2\rho}{1-\rho}\sum_{j=1,\,j\neq b}^{B} \normm{\left[ \mathbf{G}_b\mathbf{x}\right]_{\mathcal{A}_j}}_1, \, 1 \le b \le B. \vspace{-1pt}
\end{equation}
Since $\mathbf{G}_b = \operatorname{diag}(\mathbf{G}_{b1},\ldots,\mathbf{G}_{bB})$ is a diagonal matrix, its diagonal elements with entries from $\mathcal{A}_b$ are larger than $\min_{n} g_{bbn},$ and those with entries from $\mathcal{A}_j$ are smaller than $\max_{n} g_{bjn},$ which further implies
\begin{align}
	\normm{\left[ \mathbf{G}_b\mathbf{x} \right]_{\mathcal{A}_b}}_1 & \ge \big(\min_{n} g_{bbn}\big) \normv{\mathbf{x}_{\mathcal{A}_b}}_1 \ge D \normv{\mathbf{x}_{\mathcal{A}_b}}_1, \label{eq:l1-lower} \\
	\normm{\left[ \mathbf{G}_b\mathbf{x}\right]_{\mathcal{A}_j}}_1 & \le \big(\max_{n} g_{bjn}\big) \normv{\mathbf{x}_{\mathcal{A}_j}}_1, \, \forall \, j \neq b, \label{eq:l1-upper}
\end{align}
where the last inequality in \eqref{eq:l1-lower} is from \eqref{eq:large-D}.
Substituting \eqref{eq:l1-lower} and \eqref{eq:l1-upper} into \eqref{eq:reduced-main}, we get
\begin{equation}\label{eq:upper-lower-main}
	D \normv{\mathbf{x}_{\mathcal{A}_b}}_1 \le \frac{2\rho}{1-\rho}  \sum_{j=1,\,j\neq b}^{B} \big(\max_{n} g_{bjn}\big) \normv{\mathbf{x}_{\mathcal{A}_j}}_1, \, 1 \le b \le B. 
\end{equation}
In particular, consider $\bar{b} = \argmax_{b}\normv{\mathbf{x}_{\mathcal{A}_b}}_1.$
Then, we get the following contradiction:
\begin{align}\label{eq:contradictory-main}
	D \normv{\mathbf{x}_{\mathcal{A}_{\bar{b}}}}_1 & \le \frac{2\rho}{1-\rho} \sum_{j=1,\,j\neq \bar{b}}^{B} (\max_{n} g_{\bar{b}jn}) \normv{\mathbf{x}_{\mathcal{A}_j}}_1 \nonumber \\
	%& \le \frac{2\rho}{1-\rho} \sum_{j=1,\,j\neq \bar{b}}^{B} (\max_{n} g_{\bar{b}jn}) \normv{\mathbf{x}_{\mathcal{A}_{\bar{b}}}}_1 \nonumber \\
	& \le \frac{2\rho}{1-\rho} C \normv{\mathbf{x}_{\mathcal{A}_{\bar{b}}}}_1 < D \normv{\mathbf{x}_{\mathcal{A}_{\bar{b}}}}_1,
\end{align}
where the second inequality is due to \eqref{eq:less-C} and the last inequality comes from the fact $\rho < \bar{\rho} = \frac{D}{D+2C}.$ %hence $\frac{2\rho}{1-\rho} C < D.$
The contradiction in \eqref{eq:contradictory-main} implies that $\mathcal{N}\cap\mathcal{C} = \{ \mathbf{0} \}.$ This completes the proof of Theorem~\ref{theorem:scaling-law}. %\hfill $\square$

\subsection{Distribution of Estimation Error}

Next, we characterize the distribution of the estimation error $\hat{\mathbf{a}}^{(M)} - \mathbf{a}^{\circ},$ which is a generalization of \cite[Theorem 4]{chen2022phase} from the single-cell case to the multi-cell case.
\begin{theorem}\label{theo:error}
	Consider the MLE problem \eqref{eq.prob1.multi} with given $\mathbf{S},\,\{\mathbf{G}_b\},$ $\sigma_w^2,\,\mathbf{a}^{\circ}$ and $M.$
	Assume that $\mathcal{N}\cap\mathcal{C} = \{ \mathbf{0} \}$ as in Lemma~\ref{lemma:consistency} holds.
	The Fisher information matrix of problem \eqref{eq.prob1.multi} is given by \vspace{0pt}
	\begin{equation}
		\mathbf{J}(\mathbf{a}) = M \sum_{b=1}^{B} \big( \mathbf{Q}_b \odot \mathbf{Q}_b^* \big),
		\vspace{0pt}
	\end{equation}
	where
	$\mathbf{Q}_b = \mathbf{G}_b^{\frac{1}{2}} \mathbf{S}^H \big(\mathbf{S} \mathbf{G}_b\mathbf{A}\mathbf{S}^H + \sigma_w^2 \mathbf{I}  \big)^{-1} \mathbf{S} \mathbf{G}_b^{\frac{1}{2}},$ and 
	$\odot$ is the element-wise product.
	Let $\mathbf{x} \in \mathbb{R}^{BN}$ be a random vector sampled from $\mathcal{N} \big(\mathbf{0}, M \mathbf{J}(\mathbf{a}^{\circ})^{\dagger}\big),$ where $(\cdot)^\dagger$ denotes the Moore-Penrose inverse.
	Then, for each realization of $\mathbf{x},$ there exists a vector $\hat{\boldsymbol{\mu}} \in \mathbb{R}^{BN}$ which is the solution to the following QP:
	\begin{subequations}\label{eq:QP}
		\vspace{2pt}
		\begin{alignat}{2}
			&\underset{\boldsymbol\mu}{\operatorname{minimize}}    &\quad& \frac{1}{M} (\mathbf{x}-\boldsymbol\mu)^T \mathbf{J}(\mathbf{a}^{\circ})  (\mathbf{x}-\boldsymbol\mu)\\
			&\operatorname{subject\,to} &      & \boldsymbol{\mu} \in \mathcal{C}
		\end{alignat}
	\end{subequations}
	such that $\sqrt{M} \big(\hat{\mathbf{a}}^{(M)} - \mathbf{a}^{\circ}\big)$ converges in distribution to the collection of $\hat{\boldsymbol{\mu}}$'s as $M \to \infty,$ where $\mathcal{C}$ is defined in \eqref{eq:cone}.
\end{theorem}

Theorem~\ref{theo:error} shows that the estimation error $\hat{\mathbf{a}}^{(M)} - \mathbf{a}^{\circ}$ can be approximated by $\frac{1}{\sqrt{M}}\hat{\boldsymbol{\mu}}$ for a sufficiently large $M.$
It also shows that we can numerically compute the error distribution by solving QP~\eqref{eq:QP}.
We omit the proof of Theorem~\ref{theo:error} due to the space reason.

\section{Simulation Results}

Consider a multi-cell system consisting of hexagonal cells and all potential devices within each cell are uniformly distributed.
In the simulations, the radius of each cell is $500\,$m; the channel path-loss is modeled as $128.1+37.6\log_{10}(d)$ as in Assumption~\ref{assu:cell}, where $d$ is the corresponding BS-device distance in km; 
the transmit power of each device is set as $23\,$dBm, and the background noise power is $-169\,$dBm/Hz over $10\,$MHz. %\cite{3GPP}
Assume that all signature sequences of length $L$ are uniformly drawn from the sphere of radius $\sqrt{L}$ in an i.i.d. fashion as in Assumption~\ref{assu:sequence}.

\begin{figure}[t]
	\centering
	\includegraphics[width=0.42\textwidth,clip]{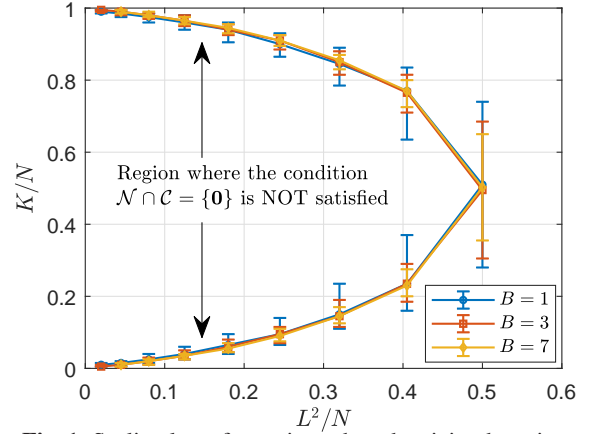}
	\vspace{-14pt}
	\caption{Scaling law of covariance based activity detection.}%for different numbers of cells
	\vspace{-10pt}
	\label{fig:phase-transition}
\end{figure}

\begin{figure}[t]
	\centering
	\includegraphics[width=0.42\textwidth,clip]{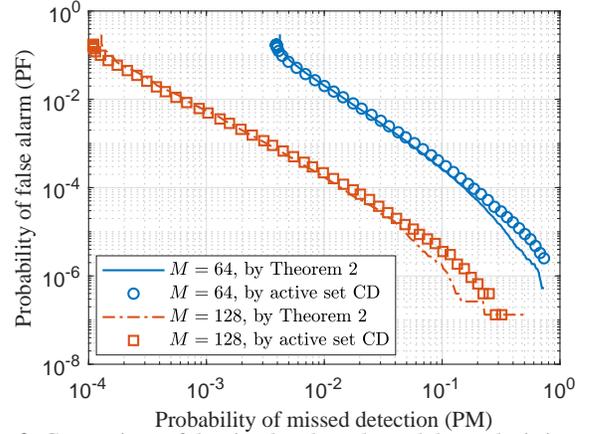}
	\vspace{-14pt}
	\caption{Comparison of the simulated results and the analysis in terms of PM and PF.}%
	%\vspace*{-0.5cm}
	\vspace{-16pt}
	\label{fig:error}
\end{figure}

We solve the linear program proposed in \cite{chen2021sparse} to numerically test the condition $\mathcal{N}\cap \mathcal{C}=\{\mathbf{0}\}$ in Lemma~\ref{lemma:consistency} under a variety of choices of $L$ and $K,$ given $N = 200$ and $B = 1, 3, 7.$
Fig.~\ref{fig:phase-transition} plots the region of $(L^2/N, K/N)$ in which the condition is satisfied or not.
The result is obtained based on $100$ random realizations of $\mathbf{S}$ and $\mathbf{a}^{\circ}$ for each given $K$ and $L.$
The error bars indicate the range beyond which either all realizations or no realization satisfy the condition.
We observe from Fig.~\ref{fig:phase-transition} that the curves with different $B$'s overlap with each other, implying that the scaling law for $\mathcal{N}\cap \mathcal{C}=\{\mathbf{0}\}$  is almost independent of $B.$
This is consistent with our analysis in Theorem~\ref{theorem:scaling-law} that the number of active devices $K$ which can be correctly detected in each cell in the multi-cell scenario depends on $B$ only through $\log B.$
We can also observe from Fig.~\ref{fig:phase-transition} that $K$ is approximately proportional to $L^2,$ which verifies the scaling law in \eqref{eq:scaling-law}.
Finally, it is worthwhile mentioning that the curves in Fig.~\ref{fig:phase-transition} are symmetric. More interpretations and discussions on the symmetry of the curves can be found in \cite{chen2021sparse}.

We validate the error estimation in Fig.~\ref{fig:error} with $B = 7,N=200,$ $ K = 20,\,L = 20,$ and $M = 64,$ or $128.$
We use the active set CD algorithm proposed in \cite{wang2021accelerating} to solve the MLE problem \eqref{eq.prob1.multi}, and compare its detection performance with the theoretical performance predicted by Theorem~\ref{theo:error}.
The probability of missed detection (PM) and probability of false alarm (PF) are traded off by choosing different values for the threshold.
We can observe from Fig.~\ref{fig:error} that the curves obtained from the active set CD algorithm match well with the theoretical ones.

\bibliographystyle{IEEEtran}
\bibliography{covariance_bib}

\begin{thebibliography}{10}
\providecommand{\url}[1]{#1}
\csname url@samestyle\endcsname
\providecommand{\newblock}{\relax}
\providecommand{\bibinfo}[2]{#2}
\providecommand{\BIBentrySTDinterwordspacing}{\spaceskip=0pt\relax}
\providecommand{\BIBentryALTinterwordstretchfactor}{4}
\providecommand{\BIBentryALTinterwordspacing}{\spaceskip=\fontdimen2\font plus
\BIBentryALTinterwordstretchfactor\fontdimen3\font minus
  \fontdimen4\font\relax}
\providecommand{\BIBforeignlanguage}[2]{{%
\expandafter\ifx\csname l@#1\endcsname\relax
\typeout{** WARNING: IEEEtran.bst: No hyphenation pattern has been}%
\typeout{** loaded for the language `#1'. Using the pattern for}%
\typeout{** the default language instead.}%
\else
\language=\csname l@#1\endcsname
\fi
#2}}
\providecommand{\BIBdecl}{\relax}
\BIBdecl

\bibitem{bockelmann2016massive}
C.~Bockelmann, N.~Pratas, H.~Nikopour, K.~Au, T.~Svensson, {\v
  C}.~Stefanovi{\'c}, P.~Popovski, and A.~Dekorsy, ``Massive machine-type
  communications in 5{G}: Physical and {MAC}-layer solutions,'' \emph{IEEE
  Commun. Mag.}, vol.~54, no.~9, pp. 59--65, Sep. 2016.

\bibitem{chen2021massive}
X.~{Chen}, D.~W.~K. {Ng}, W.~{Yu}, E.~G. {Larsson}, N.~{Al-Dhahir}, and
  R.~{Schober}, ``Massive access for {5G} and beyond,'' \emph{IEEE J. Sel.
  Areas Commun.}, vol.~39, no.~3, pp. 615--637, Mar. 2021.

\bibitem{kang2021minimum}
J.~Kang and W.~Yu, ``Minimum feedback for collision-free scheduling in massive
  random access,'' \emph{IEEE Trans. Inf. Theory}, vol.~67, no.~12, pp.
  8094--8108, Dec. 2021.

\bibitem{kang2022scheduling}
------, ``Scheduling versus contention for massive random access in massive
  {MIMO} systems,'' \emph{IEEE Trans. Commun.}, vol.~70, no.~9, pp. 5811--5824,
  Sep. 2022.

\bibitem{liu2018sparse}
L.~Liu, E.~G. Larsson, W.~Yu, P.~Popovski, {\v C}.~Stefanovi{\'c}, and
  E.~de~Carvalho, ``Sparse signal processing for grant-free massive
  connectivity: A future paradigm for random access protocols in the internet
  of things,'' \emph{IEEE Signal Process. Mag.}, vol.~35, no.~5, pp. 88--99,
  Sep. 2018.

\bibitem{senel2018grant}
K.~Senel and E.~G. Larsson, ``Grant-free massive {MTC}-enabled massive {MIMO}:
  A compressive sensing approach,'' \emph{IEEE Trans. Commun.}, vol.~66,
  no.~12, pp. 6164--6175, Dec. 2018.

\bibitem{liu2018massive}
L.~Liu and W.~Yu, ``Massive connectivity with massive {MIMO} ---{P}art {I}:
  Device activity detection and channel estimation,'' \emph{IEEE Trans. Signal
  Process.}, vol.~66, no.~11, pp. 2933--2946, Jun. 2018.

\bibitem{chen2018sparse}
Z.~Chen, F.~Sohrabi, and W.~Yu, ``Sparse activity detection for massive
  connectivity,'' \emph{IEEE Trans. Signal Process.}, vol.~66, no.~7, pp.
  1890--1904, Apr. 2018.

\bibitem{liu2021efficient}
L.~Liu and Y.-F. Liu, ``An efficient algorithm for device detection and channel
  estimation in asynchronous {IoT} systems,'' in \emph{Proc. IEEE Int. Conf.
  Acoust., Speech, Signal Process. (ICASSP)}, Jun. 2021, pp. 4815--4819.

\bibitem{haghighatshoar2018improved}
S.~Haghighatshoar, P.~Jung, and G.~Caire, ``Improved scaling law for activity
  detection in massive {MIMO} systems,'' in \emph{Proc. IEEE Int. Symp. Inf.
  Theory (ISIT)}, Jun. 2018, pp. 381--385.

\bibitem{chen2021sparse}
Z.~Chen, F.~Sohrabi, and W.~Yu, ``Sparse activity detection in multi-cell
  massive {MIMO} exploiting channel large-scale fading,'' \emph{IEEE Trans.
  Signal Process.}, vol.~69, pp. 3768--3781, Jun. 2021.

\bibitem{ganesan2021clustering}
U.~K. Ganesan, E.~Bj{\"o}rnson, and E.~G. Larsson, ``Clustering-based activity
  detection algorithms for grant-free random access in cell-free massive
  {MIMO},'' \emph{IEEE Trans. Commun.}, vol.~69, no.~11, pp. 7520--7530, Nov.
  2021.

\bibitem{fengler2021non}
A.~Fengler, S.~Haghighatshoar, P.~Jung, and G.~Caire, ``Non-bayesian activity
  detection, large-scale fading coefficient estimation, and unsourced random
  access with a massive {MIMO} receiver,'' \emph{IEEE Trans. Inf. Theory},
  vol.~67, no.~5, pp. 2925--2951, May 2021.

\bibitem{chen2022phase}
Z.~Chen, F.~Sohrabi, Y.-F. Liu, and W.~Yu, ``Phase transition analysis for
  covariance based massive random access with massive {MIMO},'' \emph{IEEE
  Trans. Inf. Theory}, vol.~68, no.~3, pp. 1696--1715, Mar. 2022.

\bibitem{chen2019covariance}
------, ``Covariance based joint activity and data detection for massive random
  access with massive {MIMO},'' in \emph{Proc. IEEE Int. Conf. Commun. (ICC)},
  May 2019, pp. 1--6.

\bibitem{shao2020cooperative}
X.~Shao, X.~Chen, D.~W.~K. Ng, C.~Zhong, and Z.~Zhang, ``Cooperative activity
  detection: {Sourced} and unsourced massive random access paradigms,''
  \emph{IEEE Trans. Signal Process.}, vol.~68, pp. 6578--6593, Nov. 2020.

\bibitem{wang2022covariance}
Z.~Wang, Y.-F. Liu, and L.~Liu, ``Covariance-based joint device activity and
  delay detection in asynchronous {mMTC},'' \emph{IEEE Signal Process. Lett.},
  vol.~29, pp. 538--542, Jan. 2022.

\bibitem{jiang2022ml}
D.~Jiang and Y.~Cui, ``{ML} and {MAP} device activity detections for grant-free
  massive access in multi-cell networks,'' \emph{IEEE Trans. Wireless Commun.},
  vol.~21, no.~6, pp. 3893--3908, Jun. 2022.

\bibitem{liu2022mle_spawc}
W.~Liu, Y.~Cui, F.~Yang, L.~Ding, and J.~Sun, ``{MLE}-based device activity
  detection for grant-free massive access under rician fading,'' in \emph{Proc.
  IEEE Workshop Signal Process. Adv. Wireless Commun. (SPAWC)}, Jul. 2022, pp.
  1--5.

\bibitem{liu2022mle_icc}
W.~Liu, Y.~Cui, F.~Yang, L.~Ding, J.~Xu, and X.~Xu, ``{MLE}-based device
  activity detection for grant-free massive access under frequency offsets,''
  in \emph{Proc. IEEE Inter. Conf. Commun. (ICC)}, May 2022, pp. 1629--1634.

\bibitem{jia2021statistical}
W.~Jiang, Y.~Jia, and Y.~Cui, ``Statistical device activity detection for
  {OFDM}-based massive grant-free access,'' \emph{IEEE Trans. Wireless
  Commun.}, Nov. 2022.

\bibitem{wang2021accelerating}
Z.~Wang, Y.-F. Liu, Z.~Chen, and W.~Yu, ``Accelerating coordinate descent via
  active set selection for device activity detection for multi-cell massive
  random access,'' in \emph{Proc. IEEE Workshop Signal Process. Adv. Wireless
  Commun. (SPAWC)}, Sep. 2021, pp. 366--370.

\bibitem{dong2022faster}
J.~Dong, J.~Zhang, Y.~Shi, and J.~H. Wang, ``Faster activity and data detection
  in massive random access: A multiarmed bandit approach,'' \emph{IEEE Internet
  of Things J.}, vol.~9, no.~15, pp. 13\,664--13\,678, Aug. 2022.

\bibitem{wang2021efficient}
Z.~Wang, Z.~Chen, Y.-F. Liu, F.~Sohrabi, and W.~Yu, ``An efficient active set
  algorithm for covariance based joint data and activity detection for massive
  random access with massive {MIMO},'' in \emph{Proc. IEEE Int. Conf. Acoust.,
  Speech, Signal Process. (ICASSP)}, Jun. 2021, pp. 4840--4844.

\bibitem{wipf2007empirical}
D.~P. Wipf and B.~D. Rao, ``An empirical {B}ayesian strategy for solving the
  simultaneous sparse approximation problem,'' \emph{IEEE Trans. Signal
  Process.}, vol.~55, no.~7, pp. 3704--3716, Jul. 2007.

\bibitem{yang2018sparse}
Z.~Yang, J.~Li, P.~Stoica, and L.~Xie, ``Sparse methods for
  direction-of-arrival estimation,'' in \emph{Academic Press Library in Signal
  Processing}.\hskip 1em plus 0.5em minus 0.4em\relax Elsevier, 2018, vol.~7,
  pp. 509--581.

\bibitem{lin2022sparsity}
Q.~Lin, Y.~Li, and Y.-C. Wu, ``Sparsity constrained joint activity and data
  detection for massive access: A difference-of-norms penalty framework,''
  \emph{IEEE Trans. Wireless Commun.}, vol.~22, no.~3, pp. 1480--1494, Mar.
  2023.

\bibitem{li2022asynchronous}
Y.~Li, Q.~Lin, Y.-F. Liu, B.~Ai, and Y.-C. Wu, ``Asynchronous activity
  detection for cell-free massive {MIMO}: From centralized to distributed
  algorithms,'' \emph{IEEE Trans. Wireless Commun.}, Oct. 2022.

\bibitem{rappaport1996wireless}
T.~S. Rappaport, \emph{Wireless Communications: Principles and Practice},
  2nd~ed.\hskip 1em plus 0.5em minus 0.4em\relax Upper Saddle River, NJ:
  Prentice-Hall, 2002.

\end{thebibliography}

\end{document}